\documentstyle[aps,prl,twocolumn,graphics,epsfig,floats]{revtex}

\newcommand{\be}{\begin{eqnarray}}
\newcommand{\ee}{\end{eqnarray}}

\def\l{\langle}
\def\r{\rangle}

\begin{document}

\title{
Natural Entanglement in Bose-Einstein Condensates}

\author{Christoph Simon}
\address {
Centre for Quantum Computation, University of Oxford, Parks Road, Oxford OX1 3PU, United Kingdom}

\date{\today}
\maketitle

\begin{abstract}
Every Bose-Einstein condensate is in a highly entangled state, as a consequence of the fact that the particles in a
condensate are distributed over space in a coherent way. It is proved that any two regions within a condensate of
finite particle number are entangled. This entanglement does not depend on the distance between the two regions.
Criteria for the presence of entanglement are derived in the context of interference experiments. For separable
states there is a trade-off between fluctuations in particle number and interference visibility.
\end{abstract}

\bigskip

Entanglement is one of the dramatic non-classical features of quantum physics. There are states of composite
quantum systems that cannot be decomposed into probabilistic combinations of product states; there is no way of
writing the density matrix of such a state $\rho$ in the form $\rho=\sum
\limits_i p_i \sigma_1^i \otimes \sigma_2^i \otimes ... \otimes \sigma_n^i$, where $\sigma_k^i$ is a state of the
$k$-th subsystem and the $p_i>0$ are probabilities. Such states $\rho$ are called inseparable or entangled. The
properties of an entangled system are not completely determined by the properties of its parts. Only entangled
states can exhibit quantum non-locality \cite{bell}. In this case there is no way of reproducing the predictions of
quantum physics with classical systems, unless there is instantaneous communication over arbitrary distances.

In the last few years, entanglement has been studied extensively in the context of quantum computation and quantum
communication \cite{ent}. Methods for creating entangled states have been found in various physical systems,
including non-linear optics, ion traps, cavity quantum electrodynamics and nuclear magnetic resonance
\cite{bouwmeester}. A fascinating open problem is whether entanglement plays a significant role in natural physical
systems. First steps towards addressing this question were made in the study of interacting spin systems
\cite{spins}. The effect of decoherence on large-scale spin entanglement has also recently been investigated in
this context \cite{simonkempe}.

Bose-Einstein condensation is another genuine quantum phenomenon which has recently attracted a large amount of
experimental and theoretical attention \cite{bec}. In a Bose-Einstein condensate (BEC), all particles (which have
to be bosons) are in the same quantum state.

Here I would like to draw attention to the fact that by its nature every BEC is a highly entangled state. Even when
there is no entanglement of the internal states \cite{sorensen}, there is always entanglement in the external
degrees of freedom. The discussion will be focussed on the entanglement between different spatial modes, but
analogous statements can be made for momentum modes. It will be shown that any two regions inside a BEC of finite
particle number are entangled with each other. The detection of this entanglement via interference will also be
discussed.

It is convenient to use the language of non-relativistic quantum field theory. The particles composing the BEC are
described by a bosonic quantum field operator $\hat{\psi}(x)$ satisfying the commutation relation
$[\hat{\psi}(x),\hat{\psi}^{\dagger}(y)]=\delta^{(3)}(x-y) \hat{\openone}$, where $x$ and $y$ are points in
three-dimensional space. The time dependence of $\hat{\psi}(x)$ is not essential for the present purpose.
Throughout the paper operators will be distinguished by hats.

A BEC of $N$ particles is described by the state $|\Psi\r=\frac{1}{\sqrt{N!}}(\hat{f}^{\dagger})^N |0\r$, where
$|0\r$ is the vacuum state and the creation operator $\hat{f}^{\dagger}$ can be written in terms of
$\hat{\psi}(x)^{\dagger}$ as $\hat{f}^{\dagger}=\int d^3 x
\, f^*(x)\hat{\psi}^{\dagger}(x)$. The mode function $f(x)$ is normalized: $\int d^3 x \, |f(x)|^2=1$, where the integrals are over
all of space. It is easy to check that $\hat{f}$ fulfills the commutation relation
$[\hat{f},\hat{f}^{\dagger}]=\hat{\openone}$.

Let us consider a certain volume inside the BEC and study its entanglement with the rest. Denoting the volume by
$A$, one can define mode operators
\be
\hat{a}=\frac{1}{\sqrt{p}}\int \limits_{x \epsilon A} d^3 x \, f(x) \hat{\psi}(x) \nonumber\\
\hat{b}=\frac{1}{\sqrt{q}}\int \limits_{x
\epsilon \bar{A}} d^3 x \, f(x) \hat{\psi}(x),
\ee
where $\bar{A}$ is the complement of $A$, $p=\int \limits_{x \epsilon A} d^3 x \, |f(x)|^2\nonumber$ and
$q=1-p=\int\limits_{x \epsilon \bar{A}} d^3 x \, |f(x)|^2$. Then one has $\hat{f}=\sqrt{p}\, \hat{a}+\sqrt{q}
\,\hat{b}$, and $\hat{a}$ and $\hat{b}$ fulfill the following commutation relations:
$[\hat{a},\hat{b}]=[\hat{a},\hat{b}^{\dagger}]=0, [\hat{a},\hat{a}^{\dagger}]=
[\hat{b},\hat{b}^{\dagger}]=\hat{\openone}$.

The state of the BEC can now be written as
\be
&&|\Psi\r=\frac{1}{\sqrt{N!}}({\sqrt{p} \hat{a}^{\dagger}+\sqrt{q} \hat{b}^{\dagger}})^N |0\r
\nonumber\\ &&=\sum \limits_{k=0}^{N} \sqrt{ {N \choose k}} \, p^{\frac{k}{2}}\,
q^{\frac{N-k}{2}} \,|k\r_a |N-k\r_b,
\label{state}
\ee
which clearly shows the entanglement between spatial modes $\hat{a}$ and $\hat{b}$. These modes can have different
physical meanings. For example, $\hat{a}$ could be the interior of the BEC and $\hat{b}$ the rest of the universe
including the BEC's boundary region; $\hat{a}$ and $\hat{b}$ could also each contain one half of the BEC.

The nature of the entanglement in (\ref{state}) is easy to see. The $N$ particles can be distributed in various
ways over the two modes $\hat{a}$ and $\hat{b}$. The state of the BEC is a coherent superposition of all these
possibilities. This clearly generalizes to more than two modes. Note that the particle numbers in $\hat{a}$ and
$\hat{b}$ have a binomial distribution.

The present kind of entanglement cannot be detected by local measurements, i.e. measurements acting on modes
$\hat{a}$ or $\hat{b}$ separately. If one is restricted to local measurements in the basis of particle number, then
(\ref{state}) is indistinguishable from a separable mixture of the various terms without any fixed phase relation
between them. One would therefore require local measurements in bases of states corresponding to superpositions of
different particle numbers, which is impossible. However, this does not mean that the entanglement in (\ref{state})
is unobservable. It can be detected by {\it joint} measurements on the two modes, for example by interference
experiments. This point will be discussed in more detail below.

Let us now study the entanglement properties of BECs in more detail. First suppose that instead of a Fock state
with exactly $N$ particles the BEC is in a coherent state of the form \cite{ssr}
\be
|\alpha\r=e^{-|\alpha|^2/2} e^{\alpha\hat{f}^{\dagger}}|0\r
=e^{-|\alpha|^2/2} e^{\alpha \int d^3 x \, f^*(x)\hat{\psi}^{\dagger}(x)}|0\r.
\ee
This can formally be rewritten as
\be
|\alpha\r=e^{-|\alpha|^2/2} \prod \limits_{x \epsilon R^3} e^{\alpha f^*(x)\hat{\psi}^{\dagger}(x)}|0\r,
\label{alpha}
\ee
where the product is over all points in space. The state (\ref{alpha}) is clearly a product state with respect to
the pointlike spatial quasi-modes $\hat{\psi}(x)$, since the vacuum does not contain any entanglement
\cite{vacuum}.

As a consequence all states that can be written as convex combinations of coherent states are also unentangled. In
particular this applies to a Poisson distribution of number states
$|n\r=\frac{(\hat{f}^{\dagger})^n}{\sqrt{n!}}|0\r$:
\be
\rho_P=\sum \limits_{n=0}^{\infty} \frac{e^{-\lambda} \lambda^n}{n!}|n\r \l n|,
\label{poisson}
\ee
which can be rewritten as a mixture of coherent states with fixed amplitude $\sqrt{\lambda}$ but random phase
\be
\rho_P=\int \limits_0^{2\pi} d\phi \, |\sqrt{\lambda}e^{i\phi}\r\l\sqrt{\lambda}e^{i\phi}|,
\ee
as can easily be shown by expanding the coherent states in the number state basis: $|\alpha\r=e^{-|\alpha|^2/2}
\sum \frac{\alpha^n}{\sqrt{n!}} |n\r$ and performing the integration over $\phi$.

Note that a state with a thermal distribution of particle numbers is also not entangled, because it can be written
as a convex combination of coherent states with a probability distribution that is a Gaussian in the coherent state
amplitude:
\be
\rho_T=(1-t) \sum \limits_{n=0}^{\infty} t^n |n\r \l n|=\int d^2 \alpha \, e^{-\beta |\alpha|^2} |\alpha\r \l \alpha|,
\label{thermal}
\ee
where the integration is over the complex plane and $\beta=(1/t)-1$. Again the identity can be shown by expressing
the coherent states in the number state basis and performing the integration.

The above results suggest that the fixedness of the particle number plays an important role in ensuring that the
state $\frac{(\hat{f}^{\dagger})^N}{\sqrt{N!}}|0\r$ is entangled. This leads naturally to the question of
subsystems. If one considers a volume containing just part of a BEC, then the particle number in this volume is no
longer fixed. Does the reduced state corresponding to such a volume still exhibit spatial entanglement of the above
type? It will now be shown that the answer is yes, and that the entanglement disappears only in the limit of a
finite subsystem of an infinitely large BEC.

Let us partition space into three regions, $A$, $B$ and $C$. For concreteness, imagine $A$ and $B$ to be well
within the region occupied by the BEC. $A$ and $B$ may be adjacent or separated. We will trace over the region $C$
and study the entanglement between the regions $A$ and $B$. Defining modes $\hat{a}$, $\hat{b}$ and $\hat{c}$
corresponding to the three regions $A$, $B$ and $C$ as above, we have
$\hat{f}^{\dagger}=\sqrt{p}\hat{a}^{\dagger}+\sqrt{q}\hat{b}^{\dagger}+\sqrt{1-p-q}\hat{c}^{\dagger}$, where $p$,
$q$ and $1-p-q$ are the probabilities for an individual particle to be found in regions $A$, $B$ and $C$
respectively. We now trace the state $|\Psi\r=\frac{(\hat{f}^{\dagger})^N}{\sqrt{N!}}|0\r$ over the $\hat{c}$-mode
in order to find the reduced density matrix for modes $\hat{a}$ and $\hat{b}$. Applying Eq. (\ref{state}) to the
split $AB-C$ it is clear that the reduced density matrix will correspond to the states
\be
\frac{(\sqrt{\frac{p}{p+q}}\hat{a}^{\dagger}+\sqrt{\frac{q}{p+q}}\hat{b}^{\dagger})^n}{\sqrt{n!}}|0\r
\label{numberst}
\ee
occuring with binomial probabilities ${N \choose n}(p+q)^n (1-p-q)^{N-n}$.

First consider the limit where $A$ and $B$ are finite-size subsystems of an infinitely large BEC. The total
particle number $N$ goes to infinity in such a way that the mean particle number in systems $A$ and $B$ together,
$\lambda=N(p+q)$, remains finite. It is easy to show that in this limit the binomial distribution tends towards a
Poisson distribution. As a consequence the reduced state is exactly of the form (\ref{poisson}) with the individual
number states of the form (\ref{numberst}). Thus in this limit there is no entanglement in the system consisting of
$A$ and $B$ only.

As long as $C$ is of finite size, there is always entanglement between $A$ and $B$. To show this, let us write the
reduced state of modes $\hat{a}$ and $\hat{b}$ explicitly:
\be
\mbox{Tr}_C |\Psi\r\l\Psi|=\sum \limits_{n=0}^{N} \sum \limits_{k,l=0}^n {N \choose n} (1-p-q)^{N-n} \sqrt{{n \choose k}
{n \choose l}} \nonumber\\(\sqrt{p})^{k+l}(\sqrt{q})^{2n-k-l} |k,n-k\r \l l,n-l|,
\label{redstate}
\ee
where $|k,n-k\r=\frac{(\hat{a}^{\dagger})^k (\hat{b}^{\dagger})^{n-k}}{\sqrt{k!(n-k)!}}|0\r$. The inseparability of
(\ref{redstate}) can be shown using the Peres-Horodecki partial transposition criterion \cite{pt}.

This can be done by first projecting the state onto suitable two-dimensional subspaces in both modes. The simplest
choice are the subspaces spanned by the states $|0\r$ and $|1\r$. The resulting state is
\be
(1-p-q)^N
|00\r\l00|+N(1-p-q)^{N-1}(p|10\r\l10|\nonumber\\+q|01\r\l01|+\sqrt{pq}|10\r\l01|+\sqrt{pq}|01\r\l10|)\nonumber\\
+\frac{N(N-1)}{2}(1-p-q)^{N-2} 2pq |11\r\l11|,
\ee
whose partial transpose has a negative eigenvalue for all values of $p$ and $q$, which is seen most easily by
calculating its determinant. Thus any two regions within a finite-size BEC are entangled with each other.

This is not the only possible choice of two-dimensional subspaces which exhibits entanglement. If one projects in
both modes onto two arbitrary Fock states $|n\r$ and $|m\r$ (the same two numbers for modes $\hat{a}$ and
$\hat{b}$), the resulting density matrix also has a non-positive partial transpose for all values of $p$ and $q$,
which can be shown in the same way as before. The amount of entanglement between the two regions $A$ and $B$
depends only on the total particle number $N$ and on the average particle numbers in the two regions, determined by
$p$ and $q$. It does not depend on the distance between the two regions. This emphasizes the spatially coherent
character of the BEC.

We have noted above that the present type of entanglement cannot be detected by separate measurements on the modes
$\hat{a}$ and $\hat{b}$. However it can be detected by joint measurements of the two modes. A (conceptually) simple
possibility is to perform an interference experiment. For simplicity consider the state
\be
|\psi\r=\frac{(\hat{a}^{\dagger}+\hat{b}^{\dagger})^N}{\sqrt{2^N N!}}|0\r,
\label{stateintf}
\ee
which has $\l\psi|\hat{a}^{\dagger}\hat{a}|\psi\r=\l\psi|\hat{b}^{\dagger}\hat{b}|\psi\r=N/2$ and a fixed total
particle number $N$. Now consider detection in the new basis of modes given by
$\hat{a}'=(\hat{a}+\hat{b})/\sqrt{2}$, $\hat{b}'=(\hat{a}-\hat{b})/\sqrt{2}$. This corresponds to a measurement
after superimposing modes $\hat{a}$ and $\hat{b}$ on a 50-50 beam splitter. The state (\ref{stateintf}) satisfies
$\l\psi|\hat{a}'^{\dagger}\hat{a}'|\psi\r=N$ and $\l\psi|\hat{b}'^{\dagger}\hat{b}'|\psi\r=0$. As a consequence of
interference, behind the beam splitter all particles are concentrated in one of the modes.

There is no separable state of modes $\hat{a}$ and $\hat{b}$ which has the same properties, namely {\it (i)} a
fixed total particle number (different from zero), i.e. $(\Delta \hat{N})^2=\l \hat{N}^2 \r-\l \hat{N} \r^2=0$,
where
$\hat{N}=\hat{a}^{\dagger}\hat{a}+\hat{b}^{\dagger}\hat{b}=\hat{a}'^{\dagger}\hat{a}'+\hat{b}'^{\dagger}\hat{b}'$,
and {\it (ii)} complete destructive interference, i.e. $\l \hat{b}'^{\dagger}\hat{b}'\r=0$, at the same time.

This can be seen in the following way. Consider a general separable state of modes $\hat{a}$ and $\hat{b}$:
\be
\rho=\sum p_i |\phi_i\r\l\phi_i|\otimes|\chi_i\r\l\chi_i|,
\label{sep}
\ee
where $|\phi_i\r$ are states of mode $\hat{a}$ and $|\chi_i\r$ are states of mode $\hat{b}$.

First assume that the particle number is fixed to be $N$. This implies $\hat{N}|\phi_i\r |\chi_i\r=N|\phi_i\r
|\chi_i\r$ for all $i$. This is possible only if $|\phi_i\r$ is an eigenstate of $\hat{a}^{\dagger}\hat{a}$ and
$|\chi_i\r$ is an eigenstate of $\hat{b}^{\dagger}\hat{b}$, i.e.$|\phi_i\r |\chi_i\r=|n_i\r|N-n_i\r$. So $\rho$
must be a convex combination of products of Fock states. But Fock states have no definite phase, therefore
destructive interference cannot occur. It is easy to see that in this case on has $\l
\hat{b}'^{\dagger}\hat{b}'
\r=\mbox{Tr}\rho \hat{b}'^{\dagger}\hat{b}'=N/2$. The particles have a 50-50 distribution behind the beam splitter,
in contrast to the result for (\ref{stateintf}).

On the other hand, one can also impose $\l \hat{b}'^{\dagger}\hat{b}'\r=0$.  This implies
\be
\frac{1}{2} \sum p_i \left( \l \hat{a}^{\dagger}\hat{a} \r_i +  \l \hat{b}^{\dagger}\hat{b} \r_i - \l \hat{a}^{\dagger}\r_i
\l \hat{b} \r_i - \l \hat{a} \r_i \l \hat{b}^{\dagger} \r_i \right)=0,
\label{nb}
\ee
where we have introduced the shorthand notation $\l \hat{a}^{\dagger}\hat{a}
\r_i=\l \phi_i|\hat{a}^{\dagger}\hat{a}|\phi_i\r, \l \hat{b}^{\dagger}\hat{b} \r_i=\l \chi_i|\hat{b}^{\dagger}\hat{b}
|\chi_i\r,\l \hat{a} \r_i=\l \phi_i|\hat{a}|\phi_i
\r, \l \hat{b} \r_i=\l \chi_i|\hat{b}|\chi_i\r$ etc.

Consider the expression in parentheses under the sum in (\ref{nb}), dropping the index $i$ for a moment. If we
define $\l \hat{a}
\r=A e^{i\alpha}$ and $\l \hat{b}\r=B e^{i\beta}$ with $A,B$ positive and $\alpha,\beta$ real, this translates into
\be
A^2+C_a+B^2+C_b-2AB\cos(\alpha-\beta).
\label{zero}
\ee
Here we have introduced the correlation functions $C_a=\l \hat{a}^{\dagger}\hat{a} \r-\l \hat{a} \r \l
\hat{a}^{\dagger}
\r$ and $C_b=\l \hat{b}^{\dagger}\hat{b} \r-\l \hat{b} \r \l \hat{b}^{\dagger} \r$, which are positive definite,
as can be seen from the Cauchy-Schwarz inequality applied to vectors $|\phi\r$ and $|v\r=\hat{a}|\phi\r$ \cite{cs}.

Therefore (\ref{zero}) is never negative, and equal to zero only if $C_a=C_b=0$, $A=B$ and $\cos(\alpha-\beta)=1$.
The Cauchy-Schwarz inequality is saturated only if the two vectors are collinear, therefore the first condition
implies $\hat{a}|\phi\r=A e^{i\alpha}|\phi\r$ and $\hat{b}|\phi\r=Be^{i\beta}|\phi\r$. Returning to Eq. (\ref{nb}),
this implies that all the states $|\phi_i\r$ and $|\chi_i\r$ have to be coherent states, with the additional
constraints $A_i=B_i$ and $\cos(\alpha_i-\beta_i)=1$ for all $i$. This implies that the particle number cannot be
fixed. For a mixture of products of coherent states
 of the form $\sum_i p_i |A_i\r \l A_i| \otimes |A_i\r \l A_i|$ one has $\l \hat{N} \r=\l \hat{a}^{\dagger}\hat{a}+
 \hat{b}^{\dagger}\hat{b} \r=2 \sum_i p_i |A_i|^2$ and $\l \hat{N}^2 \r=\l \hat{a}^{\dagger}\hat{a}\hat{a}^{\dagger}\hat{a}+
 \hat{b}^{\dagger}\hat{b}\hat{b}^{\dagger}\hat{b}+2 \hat{a}^{\dagger}\hat{a}\hat{b}^{\dagger}\hat{b}\r=
 4 \sum_i p_i |A_i|^4+ 2 \sum_i p_i |A_i|^2$, using $\hat{a}\hat{a}^{\dagger}=\hat{a}^{\dagger}\hat{a}+\hat{\openone}$.
Since $\sum_i p_i |A_i|^4$ is always greater than or equal to $(\sum_i p_i |A_i|^2)^2$, this implies that $(\Delta
\hat{N})^2 =\l \hat{N}^2 \r-\l \hat{N} \r^2 \geq \l \hat{N} \r$. If one demands perfect destructive interference, then (for
a separable state) the particle number cannot be definite.

Thus for separable states there is clearly a trade-off between the fluctuations in particle number, expressed by
$(\Delta \hat{N})^2$, and the amount of interference, expressed by $\l \hat{b}'^{\dagger}\hat{b}'\r$. If one of the
two quantities is zero, the other one is bounded away from zero. If these lower bounds are violated, the state
under consideration must be entangled. It may be possible to extend the above results, for example to derive a
region in the $((\Delta
\hat{N})^2,\l \hat{b}'^{\dagger}\hat{b}'\r)$ plane (for fixed mean particle number) where the states are definitely entangled.
This would be desirable because in a real experiment neither of the two quantities would be exactly equal to zero.
Such a more general criterion might then allow a direct experimental proof of the presence of entanglement in
Bose-Einstein condensates via interference. Note that interference between different distinct regions of a
Bose-Einstein condensate was observed in \cite{kasevich}, where the BEC was distributed over an array of
micro-traps.

The effect of non-zero temperature on the entanglement analyzed here is another topic for future research. More
generally, one may wonder whether new insights in condensate phenomena such as superfluidity can be gained from the
point of view of entanglement. Finally it should be interesting to compare the present results to the case of
condensed fermion pairs.

I would like to thank S. Bose and A. Costa for useful discussions. This work was supported by the QuComm project of
the European Union (IST-1999-10033).

\end{document}